\documentclass[journal=jpcl,manuscript=article]{achemso}
\usepackage[version=3]{mhchem}

\title{Screening and antiscreening effects in endohedral nanotubes}

\author{Pier Luigi Silvestrelli}
\email{pierluigi.silvestrelli@unipd.it}
\author{Matteo Tessarolo}
\author{Abdolvahab Seif}
\author{Alberto Ambrosetti}
\affiliation{Dipartimento di Fisica e Astronomia, Universit\`a degli 
Studi di Padova, 35131 Padova, Italy}

\date{\today}

\begin{document}

\newpage

\begin{abstract}
Recently we investigated from first principles screening properties in systems 
where small molecules, characterized by a finite electronic dipole moment,
are encapsulated into different nanocages.
The most relevant result was the observation of an
antiscreening effect in alkali-halide nanocages characterized by ionic bonds:
in fact, due to the relative displacement 
of positive and negative ions, induced by the dipole moment of the 
encapsulated molecule, these cages act as dipole-field amplifiers,
differently from what observed in carbon fullerene nanocages, which
exhibit instead a pronounced screening effect.
Here we extend the study to another class of nanostructures,
the nanotubes. 
Using first-principles techniques based on the Density Functional Theory,
we studied the properties of endohedral nanotubes obtained by 
encapsulation of a water molecule or a linear HF molecule. 
A detailed analysis of the effective dipole moment of the
complexes and of the electronic charge distribution suggests that
screening effects crucially depend not only on the nature of the
intramolecular bonds but also on the size and the shape
of the nanotubes, and on the specific encapsulated molecule.
As observed in endohedral nanocages, screening is maximum
in covalent-bond carbon nanotubes, while it is reduced in partially-ionic
nanotubes and an antiscreening effect is observed in some ionic nanotubes.
However in this case the scenario is more complex than in corresponding 
ionic nanocages.
In fact the specific geometric structure of alkali-halide nanotubes 
turns out to be crucial for determining the
screening/antiscreening behavior: while nanotubes
with {\it octagonal} transversal section can exhibit an
antiscreening effect, which quantitatively depends on the number
of layers in the longitudinal direction, instead nanotubes with
{\it dodecagonal} section are always characterized by a reduction of the total
dipole moment, so that a screening behavior is observed.
Our results therefore show that, even in nanotube structures, in principle 
one can tune the dipole moment and generate electrostatic fields at the 
nanoscale without the aid of external potentials.
\end{abstract}



\maketitle

\section{Introduction}
Since the discovery\cite{Kroto} of Buckminsterfullerene (C$_{60}$)
this complex has
received intense study, also considering that, although C$_{60}$ is the most
stable and the most common naturally occurring fullerene, many other
cage-like nanostructures have been obtained and can be hypothesized,
by both considering different
numbers of carbon atoms and also replacing carbons with other atoms.
Interestingly, by high-energy collisions of ionized fullerene species, harsh
conditions of high temperature and pressure, electric arc, or by organic
synthesis methods (``molecular surgery''), it is nowadays possible to 
produce C$_{60}$ endohedral complexes with metal ions, noble gases,
and small molecules, such as H$_2$, N$_2$, H$_2$O, and CH$_4$ 
(the first organic molecule to be 
encapsulated)\cite{Akasaka,Suetsuna,Komatsu,Kurotobi,Bloodworth,Jaworski,Sinha}.
Such recent achievements in the synthesis of endohedral fullerene complexes
have stimulated many experimental and theoretical investigations since 
the cavity inside fullerenes provides a unique environment for the study of 
isolated atoms and molecules. Moreover, these systems represent ideal models 
to study how confinement effects can induce changes in structural and 
electronic properties of small molecular species and also provide a 
possible way to alter the properties of the otherwise rather inert fullerenes.
In particular, Kurotobi and Murata developed a synthetic route to surgically 
insert a single water molecule into the most common fullerene 
C$_{60}$\cite{Kurotobi}. 
In this case the water molecule, with its relatively
large dipole moment (1.8 D), is expected to polarize the 
symmetric non-polar C$_{60}$ cage. 
This was confirmed by theoretical first-principles 
studies\cite{Ramachandran,Yagi,Ensing} showing that the dipole moment of 
H$_2$O@C$_{60}$ is much lower (about 0.5 D) than that for the isolated water, 
thus suggesting that a substantial counteracting dipole moment is induced 
in the C$_{60}$ cage, which considerably screens the electric field produced by 
the dipole moment of the encapsulated water molecule. 

In a recent paper\cite{psil23} we have investigated 
endohedral nanocages, characterized by different interatomic interactions, 
also considerning other encapsulated molecules than water.
Our calculations of binding and
electronic properties, and detailed analysis of the effective 
dipole moment of the complexes and the electronic charge distribution elucidated
the encapsulation effects and suggest that
the screening phenomenon crucially depends on the nature of the
intramolecular bonds of the cage: screening is maximum
in covalent-bond carbon nanocages, is reduced in partially-ionic
ones, while in the case of the ionic-bond nanocages 
(i.e. Li$_{36}$F$_{36}$) an {\it antiscreening} effect is observed. 
Hence, the latter systems surprisingly act as dipole-field amplifiers.

Here we extend the study to different nanotubes:
metal carbon nanotubes, partially ionic nanotubes 
(with alternate B and N, or Be and O, or Ga and N atoms), 
and ionic, alkali-halide nanotubes. 
Alkali-halide nanotubes are currently investigated both experimentally 
and theoretically, particularly focusing on LiF nanotubes because these 
nanostructures can be obtaining experimentally by the impact of fast ions 
on a LiF polycrystal.\cite{Lima,Mussi,KimY}
On the basis of a relative-stability analysis,\cite{Lima} 
the most stable LiF nanotube structures (and the most suitable for being 
experimentally realised) turns out to be characterized by {\it hexagonal}
and {\it octagonal} bases: these structures exhibit a stability similar or
higher than that of corresponding cubic clusters.\cite{Lima}
In particular, first-principles simulations
showed\cite{Lima} that a LiF nanotube with octagonal base and 7 layers
is stable at room temperature and atmospheric pressure. 
Therefore, since nanotubes with hexagonal
section are expected to be too small to allow encapsulation of small molecules,
we consider, as a reference alkali-halide system the Li$_{24}$F$_{24}$ 
nanotube made up of
24 LiF pairs, with {\it octagonal} base and six layers, although we 
have subsequently investigated also LiF nanotubes with 
{\it dodecagonal} section. 

After a preliminary evaluation of the
thermal stability of several nanotubes, characterized by intermolecular 
bonds of different nature, we focus on the properties of the most stable
structures. In particular, by considering different binary combinations of 
alkali metals and halides, we confirm that structures made up of lithium and 
fluorine atoms are the most stable alkali-halide nanotubes.
Previous first-principles calculations showed indeed
the high stability of alkali-halide LiF nanotubes\cite{Lima} and stable 
nanotube structures were also found for NaCl and KBr.

Using first-principles techniques based on the Density Functional Theory (DFT),
we then studied the properties of a number of endohedral nanotubes obtained by 
encapsulation of a water molecule or a linear HF molecule. 
Similarly to nanocages, also in nanotubes the screening effects crucially 
depend on the nature of the intramolecular bonds: screening is maximum in 
covalent-bond carbon nanotubes, while it is reduced in partially-ionic 
nanotubes and an antiscreening effect is observed in some ionic nanotubes.
However in this case the scenario is more complex since screening effects 
also depend on the size and the shape of the nanotubes. 
In particular, the specific geometric
structure of alkali-halide, ionic nanotubes turns out to be crucial 
for determining the screening/antiscreening behavior. In fact, while nanotubes 
with {\it octagonal} transversal section can exhibit an
antiscreening effect, which quantitatively depends on the number
of layers in the longitudinal direction, instead nanotubes with
{\it dodecagonal} section are always characterized by a reduction of the total
dipole moment, so that a screening behavior is observed.
The antiscreening effect in octagonal nanotubes is particularly
pronounced for the encapsulated water molecule or the HF-H$_2$O chained complex,
characterized by the formation of Hydrogen bonds.
A possible explanation of this intriguing behavior is proposed on the basis
of a detailed structural and electronic analysis.

\section{Method}
We denote the investigated endohedral complexes as X@Y, where X=HF or H$_2$O
or HF-H$_2$O, while Y = C$_n$ ($n=$ 96,120) or B$_{48}$N$_{48}$ or 
Be$_{48}$O$_{48}$ or Ga$_{48}$N$_{48}$ or Li$_{n}$F$_{n}$ 
($n=$ 20,24,28,30,36,42).
Nanotubes are characterized not only by the number and type of constituent 
atoms, but also by their transversal section and the number
of layers in the longitudinal direction. For instance, one of the most 
interesting system in this study is represented by 
the H$_2$O@Li$_{24}$F$_{24}${\it oct-6} structure (see Figure 1), namely a 
lithium fluoride octagonal nanotube made up of six layers encapsulating a 
water molecule. 
Smaller alkali-halide nanotubes, such as Li$_{18}$F$_{18}$, turn out to be 
too small to encapsulate even the simple HF molecule: 
in fact the HF@Li$_{18}$F$_{18}${\it oct-6}
endohedral complex is unstable and the nanotube structure is deformed.
The same happens if one tries to encapsulate larger molecules, such as LiF,
inside the Li$_{24}$F$_{24}${\it oct-6} nanotube.
In the case of carbon nanotubes we considered both a C$_{96}$ (4,4)
and a C$_{120}$ (5,5) metallic, ``armchair'' structure.

\begin{figure}
\centering
\includegraphics[width=12cm]{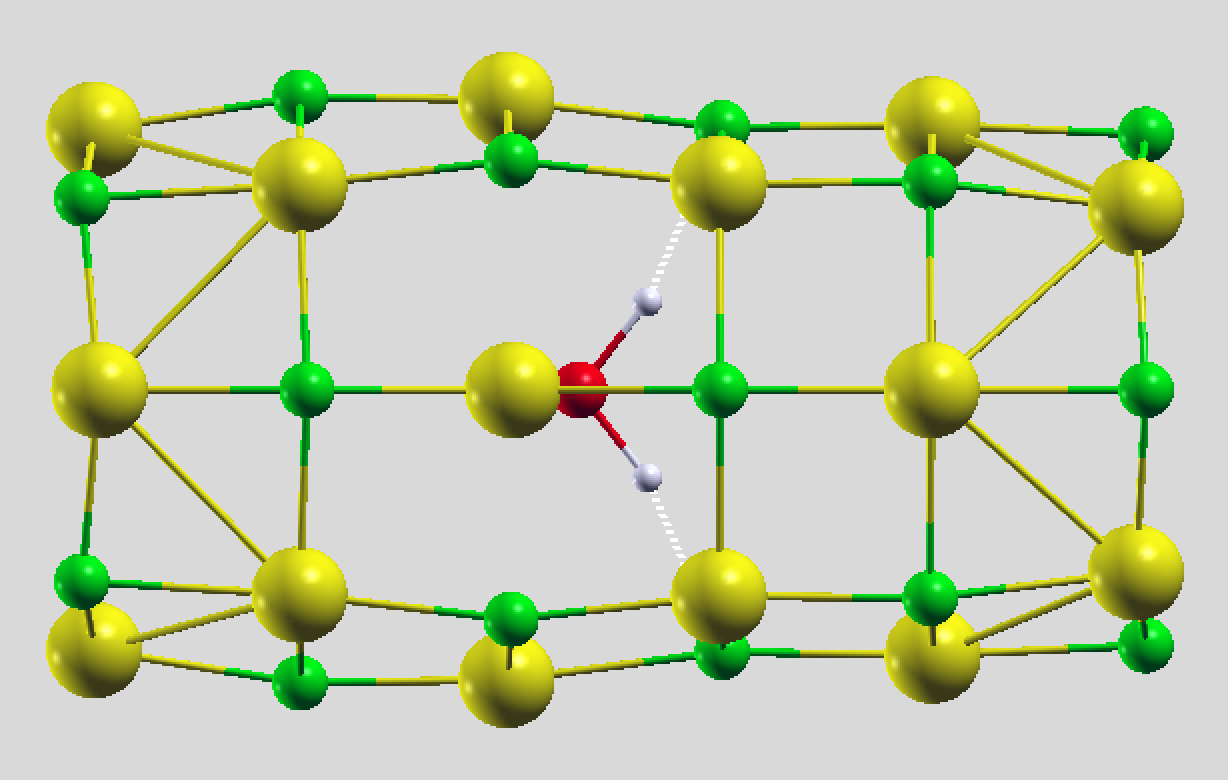} \includegraphics[width=7cm]{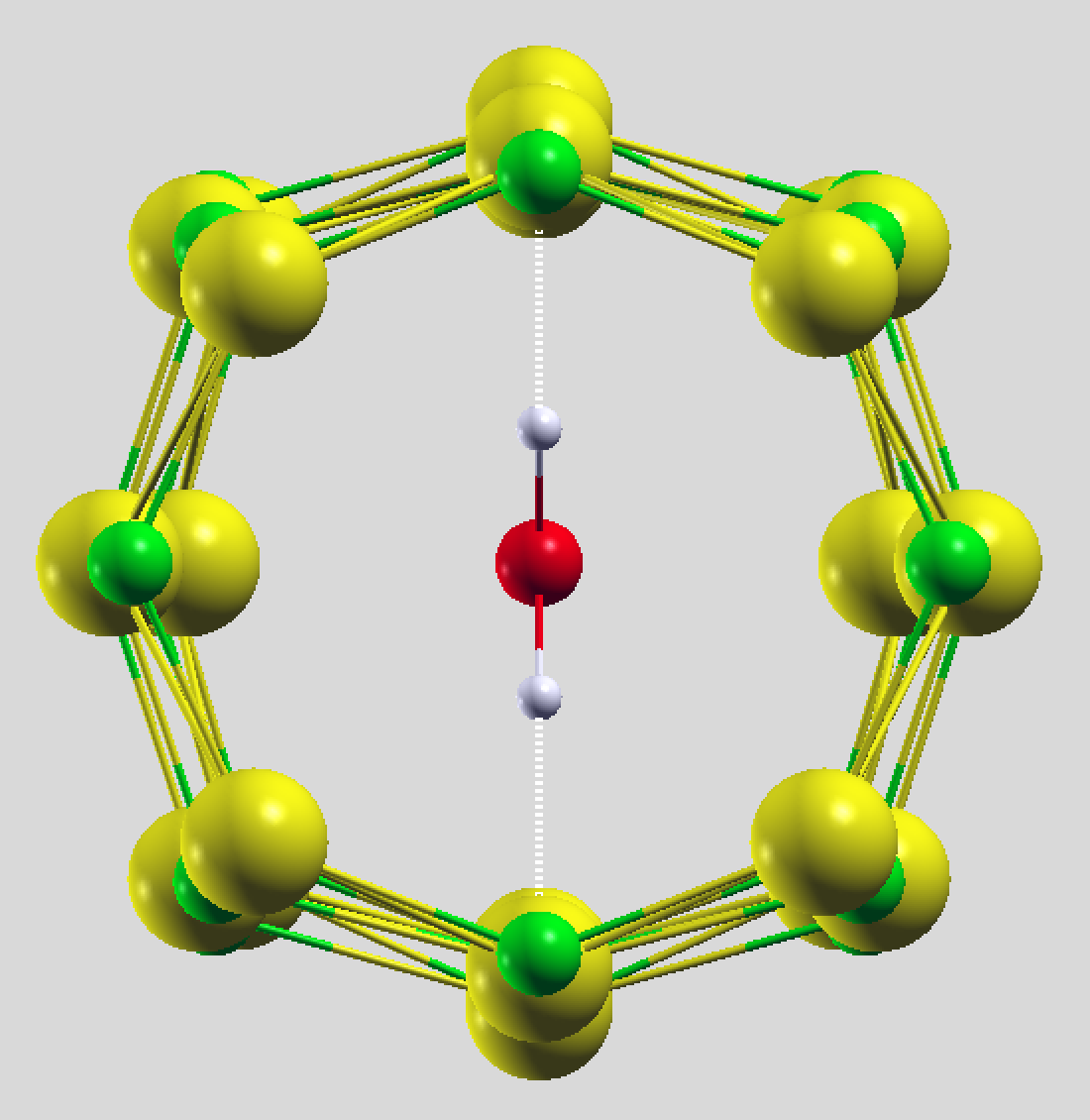} 
\caption{H$_2$O@Li$_{24}$F$_{24}${\it oct-6} endohedral nanostructure.
Longitudinal (upper part) and transversal (lower part) view.
Yellow, green, red, and white balls represent Li, F, O ,and H atoms,
respectively. White dashed lines indicate Hydrogen bonds.} 
\label{fig1}
\end{figure}

Our first-principles simulations have been performed
with the Quantum-ESPRESSO {\it ab initio} package\cite{ESPRESSO}, 
within the framework of the DFT.
In order to simulate endohedral structures made up of isolated nanotubes,
we adopted a tetragonal simulation supercell with periodic boundary conditions. 
The axis of the nanotube coincides with the $z$ direction of the supercell.
This lattice structure was chosen to optimize the ratio between the volume 
occupied by the structure and the total volume of the supercell. 
The size of supercell has been the subject of a detailed convergence study 
to find the best compromise between the computational effort and the 
accuracy of the results, considering that it must be
sufficiently large to avoid significant spurious interactions due 
to periodic replicas. For our calculations the ratio of $c$, the dimension 
along $z$, and $a$, the dimension along $x$ and $y$, of the supercell was kept
constant and equal to 2 to match the typical size of the nanotubes, with $a$
ranging from 12 to 15 \AA. 

The sampling of the Brillouin Zone was restricted to the $\Gamma$-point 
only, which is appropriate for the study of isolated nanostructures.
Electron-ion interactions were described using ultrasoft
pseudopotentials and the wavefunctions were expanded in a plane-wave basis
set with an energy cutoff of 70 Ry.
Since van der Waals (vdW) forces are expected to play an important role 
in the interaction of an encapsulated molecule with the surrounding 
nanotube\cite{Ensing,psil23}, the calculations were performed by adopting the
rVV10 DFT functional\cite{Sabatini} (the revised, more efficient
version of the original VV10 scheme\cite{Vydrov}), where
vdW effects are included by introducing an explicitly nonlocal correlation
functional.
rVV10 has been found to perform well in many systems and phenomena
where vdW effects are relevant, including several adsorption
processes\cite{Sabatini,psil15,psil16}.
In order to elucidate the specific effects of vdW corrections, in some 
cases binding energies
were also computed by replacing the rVV10 DFT functional with the PBE\cite{PBE} 
one, which is a functional unable to properly describe vdW effects. 
The approach used to accurately describe the screening/antiscreening effect
is similar to that adopted in ref.\cite{psil23}. It basically consists in
the evaluation of the dipole moment from the valence electron density 
distribution and the nuclei positions, in the analysis of the differential 
charge density, and, for ionic nanostructures, in
the calculation of the displacements of positive and negative ions of the
nanotube upon encapsulation of a small molecule. 
The electronic dipole moment of the systems is computed as:

\begin{equation}
\mu = -e\int d{\bf r}\, {\bf r}\, n({\bf r}) + \sum_i^{N} Z_i e {\bf R}_i \;,
\end{equation}
where $-e$ is the electron charge, $n({\bf r})$ the electronic
number density, and $Z_i$ and ${\bf R}_i$ are the valence and spatial 
coordinate of the $i$-th ion of the system, respectively.

\section{Results}
Preliminary stability tests for several nanotubes were performed 
considering both their cohesive energy and thermal stability.
The cohesive energy (per atom) is defined as:
\begin{equation}
E_c = \left( E - \sum_i^{N} E_i \right)/N  \;,
\end{equation}
where $E$ is the total energy of the system, $E_i$ is the energy of the
isolated $i$-th atom, and $N$ is the total number of
atoms of the system.

\begin{table}
\caption{Diameter $d$, cohesive energy $E_c$, energy gap $E_g$, and 
Lindemann number $L$ (at 300 K) of different nanotubes (see text). 
{\it oct-6} and {\it dod-6} denote octagonal or dodecagonal transversal section 
with 6 layers. For comparison, data relative to selected nanocage structures
are also reported.\cite{psil23}}
\begin{center}
\begin{tabular}{|c|c|c|c|c|}
\hline
system & $d$ (\AA) & $E_c$ (eV) & $E_g$ (eV) & $L$ \\ 
\hline
C$_{96}$ (4,4)                & 5.53 & -7.98 & 0.14 & 0.019 \\ 
C$_{120}$ (5,5)               & 6.83 & -8.06 & 0.05 & 0.018 \\ 
B$_{48}$N$_{48}$ (4,4)        & 5.52 & -6.36 & 3.92 & 0.019 \\
Be$_{48}$O$_{48}$ (4,4)       & 5.71 & -6.16 & 4.87 & 0.025 \\
Ga$_{48}$N$_{48}$ (4,4)       & 7.05 & -3.59 & 1.50 & 0.021 \\
Li$_{24}$F$_{24}${\it oct-6}  & 4.90 & -4.23 & 6.40 & 0.053 \\  
Li$_{36}$F$_{36}${\it dod-6}  & 7.38 & -4.23 & 6.07 & 0.064 \\
\hline
C$_{60}$  nanocage            & 7.10 & -7.92 & 1.65 & 0.017 \\
B$_{36}$N$_{36}$ nanocage     & 8.70 & -7.65 & 4.48 & 0.019 \\
Be$_{36}$O$_{36}$ nanocage    & 8.03 & -6.09 & 4.89 & 0.024 \\
Si$_{60}$ nanocage            &11.67 & -3.86 & 0.39 & 0.100 \\
Li$_{36}$F$_{36}$ nanocage    & 9.36 & -4.16 & 5.86 & 0.050 \\     
\hline
\end{tabular}                                                
\end{center}
\label{table1}                                  
\end{table}

As can be seen in Table I, the cohesive energies and energy gaps of 
the considered nanotubes are similar to those of nanocages made up of 
the same constituent atoms. In particular,
the cohesive energy of the LiF nanotubes is slightly larger (in absolute 
value) than that of the Li$_{36}$F$_{36}$ nanocage investigated 
in ref.\cite{psil23}.

As far as the thermal stability is concerned,
the Lindemann parameter, widely used for determining 
the melting temperatures of bulk solids is not appropriate for application
to nanotubes.\cite{ZhangK,Dietel}
Instead, a modified Lindemann number $L$, based on nearest-neighbor distance 
fluctuations, turns out to be effective:\cite{Dietel}

\begin{equation}
L = {1 \over {N_1}} \sum_{i,j=1,N_1} {\sqrt{\left<r_{ij}^2\right>-\left<r_{ij}\right>^2} \over {\left<r_{ij}\right>}}
\end{equation}

where $N_1$ is the number of all nearest-neighbor atom pairs, $r_{ij}$ 
is the modulus of the difference vector between nearest-neighbor atoms, and 
$\left<...\right>$ denotes average over the Molecular Dynamics (MD) 
simulation steps. In the present study
MD simulations were performed, at different temperatures, using the Verlet 
algorithm to integrate the equations of motion, with a time step of 
0.968 fs and a total simulation time of about 9.7 ps.
The temperature was controlled by the velocity scaling method.
Below the melting temperature $L$ is expected to increase slowly and linearly 
with temperature, due to the stable constraint of the atomic potentials and the 
linear increase in the kinetic energy, just reflecting atomic thermal vibrations
around the original positions; instead above the melting temperature
$L$ is more sensitive to the temperature change and deviate from linear 
behaviour by exhibiting an abrupt increase.\cite{ZhangK,Dietel} 
At the melting temperature $L$ should be characterized by a value of the order
of 0.050.\cite{ZhangK,Dietel}

As expected, estimated $L$ values of Table I indicate that the thermal 
stability of carbon and partially ionic nanotubes at room temperature is 
higher than that of ionic nanotubes, in line with the behavior observed in 
the cohesive energy and in the corresponding nanocages.  
By comparing $L$ in LiF nanotubes one can see that the thermal stability
of the Li$_{24}$F$_{24}${\it oct-6} is slightly higher than that of
Li$_{36}$F$_{36}${\it dod-6} although the cohesive energies of the 
two systems are essentially equal.

In Table II we report the basic properties of endohedral complexes made 
up of H$_2$O and HF molecules encapsulated inside the selected nanotubes.
Extensive calculations indicate that in any case the most favored configuration
is typically represented (see, for example, H$_2$O@Li$_{24}$F$_{24}${\it oct-6} 
in Figure 1) by the encapsulated molecule located in a central positions 
inside the nanotubes with its dipole moment parallel to the axis of 
the nanostructure.
In some endohedral structures (H$_2$O@B$_{48}$N$_{48}$ (4,4), 
H$_2$O@Be$_{48}$O$_{48}$ (4,4), H$_2$O@Li$_{24}$F$_{24}${\it oct-6},
and H$_2$O@Li$_{36}$F$_{36}${\it dod-6}) the encapsulated water molecule
forms 2 Hydrogen bonds with, respectively, the N, O, F atoms of the 
surrounding nanotube.

The binding energies (in meV, 1 kJ/mol=10.36 meV, 
1 kcal/mol=43.36 meV) of endohedral systems are
computed as the difference between the total energy
of the X@Y complex and the sum of the total energies of the constituent
parts X and Y:
\begin{equation}
E_{\rm bind} = E({\rm{X@Y}}) - E({\rm{X}}) - E({\rm{Y}}) \;.
\end{equation}
In some cases we also report the binding energies obtained by replacing 
the rVV10 DFT functional with the PBE one, so without properly taking 
vdW interactions into account.
Note that in the present systems zero-point energy (ZPE) effects 
are expected to 
be small: in fact Dolgonos and Peslherbe\cite{Dolgonos2014} verified that
the stability of endohedral complexes is not considerably affected by ZPE 
correction of the interaction energies, which does not exceed 10\%. 
Clearly a negative value of the binding energy 
indicates that the molecule energetically prefers to be encapsulated 
inside the nanotube
rather than being isolated, although an energy barrier may have to be 
overcome to penetrate the nanotube (see detailed discussion below).
In a few cases (H$_2$O@C$_{96}$ (4,4), HF@C$_{96}$ (4,4), 
HF@B$_{48}$N$_{48}$ (4,4)) the computed binding energy is positive,
thus suggesting that the configuration with the encapsulated
molecule is unfavored. Note that, in the case of carbon nanotubes,
while the binding energy is positive for the molecules inside
C$_{96}$ (4,4) with a diameter of $d=5.53$ \AA, it becomes negative 
considering the larger C$_{120}$ (5,5) nanotube (diameter of $d=6.83$ \AA).
As can be seen, similarly to what found in endohedral nanocages,\cite{psil23}
the binding energy of the encapsulated molecules is mostly due to 
vdW interactions;
in fact using the PBE functional (unable to reproduce vdW effects) the binding
energy is much reduced (in absolute value) and becomes even positive in 
H$_2$O@Li$_{24}$F$_{24}${\it oct-6} and HF@Li$_{24}$F$_{24}${\it oct-6}.

All the considered nanotubes, in their optimized, isolated structure, are 
characterized by a negligible total dipole moment (in all cases not larger
than 0.1 D, with the exception of the Be$_{48}$O$_{48}$ (4,4) nanotube
with a moderate dipole moment of 0.26 D along the longitudinal axis). 
The scenario changes when a small molecule with a finite electronic 
dipole moment is encapsulated into the nanotubes, as can be seen in 
Table II where we report both the total dipole moment of the endohedral 
complexes and the change (in absolute value and in percentage) of the 
dipole moment with respect to that of the isolated H$_2$O or HF molecule, 
due to the presence of the surrounding nanotube. 
Remarkably, in the investigated C nanotubes the screening effect
is almost total (more than 90\%), being even more pronounced than that observed
in C nanocages.\cite{psil23}
In partially-ionic nanotubes the screening effect is instead intermediate and 
comparable to that found in the corresponding nanocages,\cite{psil23} ranging
from 16 to 63\%.

Interestingly, in the case of the H$_2$O@Li$_{24}$F$_{24}${\it oct-6} and 
HF@Li$_{24}$F$_{24}${\it oct-6} 
endohedral complexes an evident {\it antiscreening} effect is detected,
with a noticeable dipole amplification.
This effect is comparable to that observed in the corresponding LiF 
alkali-halide nanocage\cite{psil23} for the encapsulated HF molecule 
(+16\% vs. +20\%) and even significantly stronger for the encapsulated 
water molecule (+35\% vs. +18\%).  
In order to find to what extent the rearrangement of the 
Li$_{24}$F$_{24}${\it oct-6} 
nanotube atoms can induce the dipole amplification we performed constrained 
relaxations for both H$_2$O@Li$_{24}$F$_{24}${\it oct-6} and 
HF@Li$_{24}$F$_{24}${\it oct-6}. If only the $z$ 
coordinates (those along the nanotube axis and the molecule dipole moment) 
of the nanotube atoms are allowed to
relax, the optimized configurations are energetically unstable (positive
binding energy) and a moderate dipole reduction of the endohedral complexes
is observed; so this axial relaxation cannot be the source of the antiscreening 
effect, at difference from what found in alkali-halide nanocages.\cite{psil23}
Instead configuration stability and antiscreening effect are recovered by 
just allowing the radial relaxation of the Li$_{24}$F$_{24}${\it oct-6} 
nanotube (i.e. relaxing $x$ and $y$ atomic coordinates only). 
As a consequence of the molecule encapsulation the 
Li$_{24}$F$_{24}${\it oct-6} nanotube undergoes a
global {\it radial expansion}, more pronounced in 
H$_2$O@Li$_{24}$F$_{24}${\it oct-6}
than in HF@Li$_{24}$F$_{24}${\it oct-6}: in particular, 
the average radial expansion
of the Li and F atoms is 3.6 and 2.2 m\AA\ in HF@Li$_{24}$F$_{24}${\it oct-6} 
and 29.0 and 26.3 m\AA\ in H$_2$O@Li$_{24}$F$_{24}${\it oct-6}, this 
quantitative difference being explained by the larger size of the 
H$_2$O molecule than of HF.
Also note that the water molecule forms 2 Hydrogen bonds with 2 F atoms
of the Li$_{24}$F$_{24}${\it oct-6} nanotube (see Figure 1), which further 
contributes to the dipole-moment amplification and explains the stronger 
binding of the H$_2$O molecule than of HF.
In alkali-halide LiF complexes, due to charge redistribution, the Li and F
atoms are positively and negatively charged, so they can be considered actually
as Li$^+$ and F$^-$ ions. So, it is just this radial ion rearrangement that 
produces the antiscreening effect. 

However, the size of the alkali-halide nanotube plays a crucial role in
determining the screening properties of the alkali-halide nanotubes.
In fact, in the case of the H$_2$O@Li$_{36}$F$_{36}${\it dod-6} and 
HF@Li$_{36}$F$_{36}${\it dod-6}
endohedral complexes, the scenario is different.
From one hand the binding energies are again quite similar for H$_2$O and
HF but much larger
(in absolute value) than in the corresponding structures with 
Li$_{24}$F$_{24}${\it oct-6}
nanotubes; H$_2$O@Li$_{36}$F$_{36}${\it dod-6} and 
HF@Li$_{36}$F$_{36}${\it dod-6} 
are stable structures even with the PBE functional, so here vdW effects are 
important but slightly less crucial for structure stabilization.
More importantly, here a clear {\it screening} effect is found, 
quantitatively similar to that observed (see Table II) in 
B$_{48}$N$_{48}$ (4,4), Be$_{48}$O$_{48}$ (4,4), and Ga$_{48}$N$_{48}$ (4,4) 
partially ionic nanotubes.
In fact there is a significant dipole reduction, which is more pronounced
in H$_2$O@Li$_{36}$F$_{36}${\it dod-6}.
Differently from what happens for the Li$_{24}$F$_{24}${\it oct-6} nanotube, 
in this case a significant radial contraction is observed upon encapsulation 
of a small molecule: the average radial contraction of the Li and F atoms is 
192.3 and 198.3 m\AA\ in HF@Li$_{36}$F$_{36}${\it dod-6} and
32.4 and 34.6 m\AA\ in H$_2$O@Li$_{36}$F$_{36}${\it dod-6} (see Figure 2).  
Here the screening effect is recovered by just relaxing the $z$ coordinates
of the nanotube atoms.

\begin{figure}
\centering
\includegraphics[width=12cm]{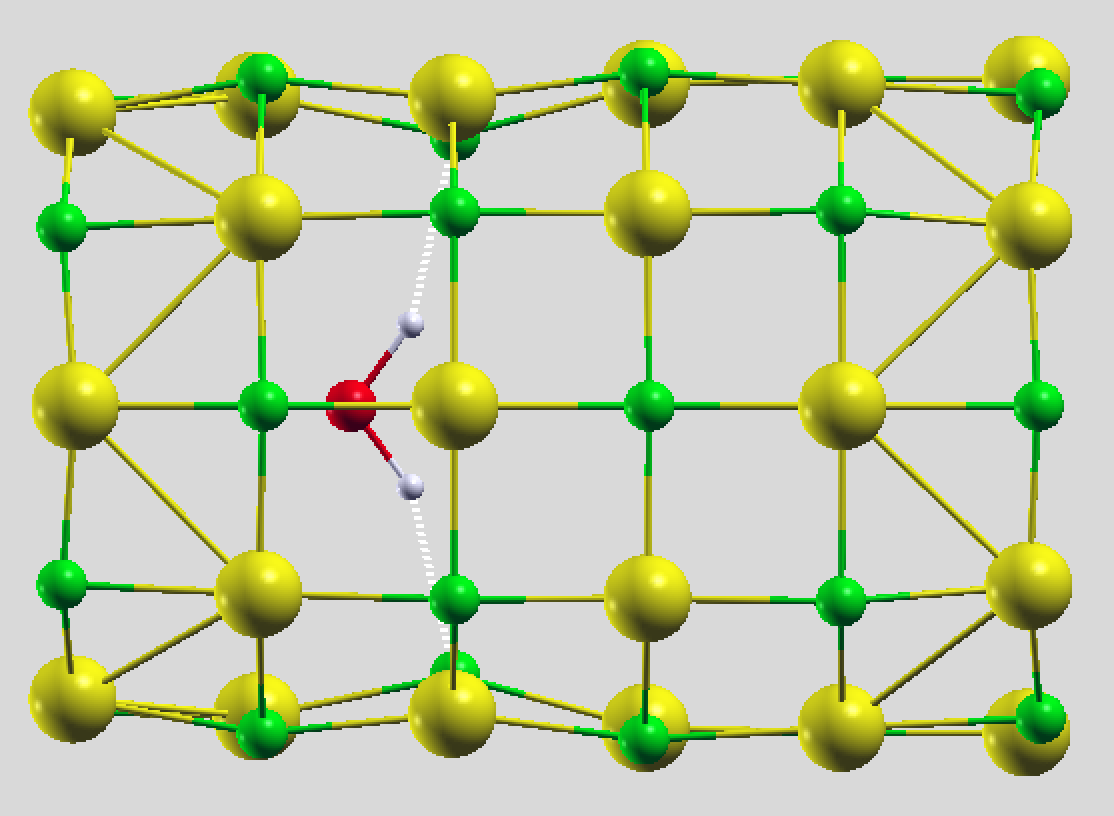} \includegraphics[width=7cm]{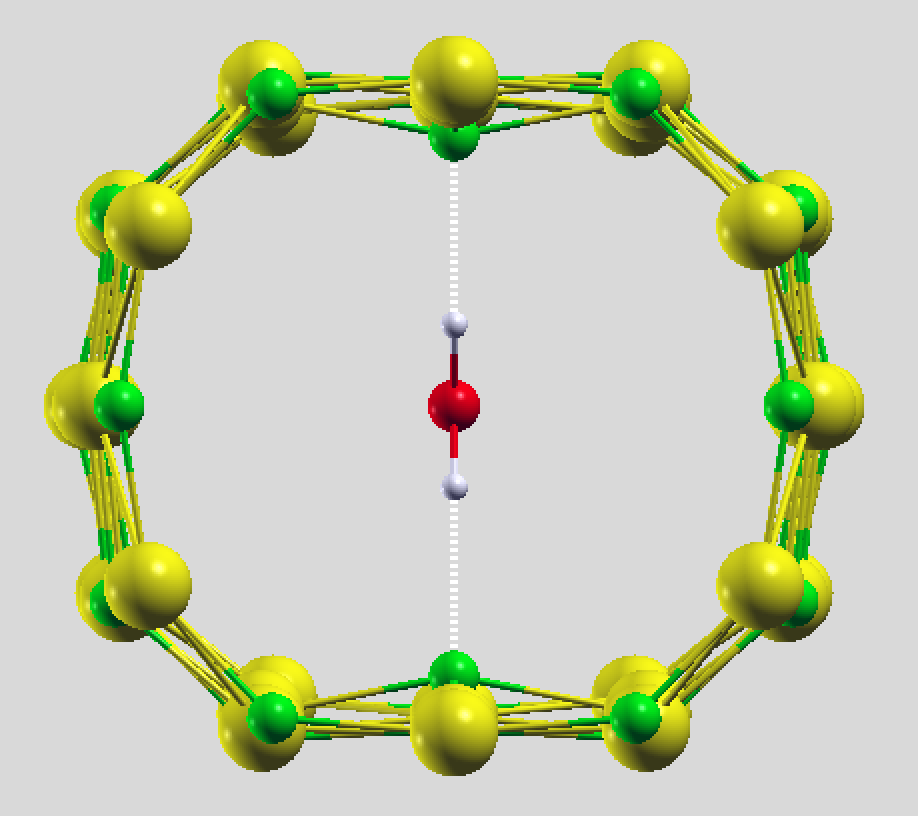} 
\caption{H$_2$O@Li$_{36}$F$_{36}${\it dod-6} endohedral nanostructure.
Longitudinal (upper part) and transversal (lower part) view.
Yellow, green, red, and white balls represent Li, F, O ,and H atoms,
respectively. White dashed lines indicate Hydrogen bonds.} 
\label{fig2}
\end{figure}

So, basically, radial atomic displacements appear to be the main source
of the antiscreening effect in Li$_{24}$F$_{24}${\it oct-6}, while 
longitudinal ones (along the $z$ coordinate) are the main source of 
screening in Li$_{36}$F$_{36}${\it dod-6}.
We also note that, upon encapsulation, the Li$_{36}$F$_{36}${\it dod-6} 
nanotube is significantly more deformed than Li$_{24}$F$_{24}${\it oct-6}
(see Figure 2). 
This effect can be made quantitative by
estimating the {\it eccentricity} of the nanostructures, which can be 
defined as

\begin{equation}
e = {\sqrt{1-(R_2/R_1)^2}}
\end{equation}

where $R_1$ and $R_2$ are the largest and the smallest nanotube radius, 
respectively.
The computed $e$ values are 0.08 and 0.27 for HF@Li$_{24}$F$_{24}${\it oct-6} 
and H$_2$O@Li$_{24}$F$_{24}${\it oct-6}, to be compared to the $e$ values 
of 0.65 and 0.80 for
HF@Li$_{36}$F$_{36}${\it dod-6} and H$_2$O@Li$_{36}$F$_{36}${\it dod-6}.

In order to better elucidate the mechanisms underlying the dipole-moment 
variations in endohedral nanotubes, a further analysis is performed
by focusing on the electron-density variations. 
In Figures 3-5 we show the 
changes in the electron distribution, for the H$_2$O@C$_{120}$ (5,5), 
H$_2$O@Li$_{24}$F$_{24}${\it oct-6}, and H$_2$O@Li$_{36}$F$_{36}${\it dod-6} 
systems, resulting from the encapsulation processes. 
The plotted differential charge density,
$\delta\rho$, is defined as the difference between the total electron
density of the whole system and the superposition of the densities of the 
separated fragments (H$_2$O molecule and nanotube), keeping the same 
geometrical structure and atomic positions that these fragments have within 
the whole optimized system. 

\begin{figure}
\centerline{
\includegraphics[width=13cm]{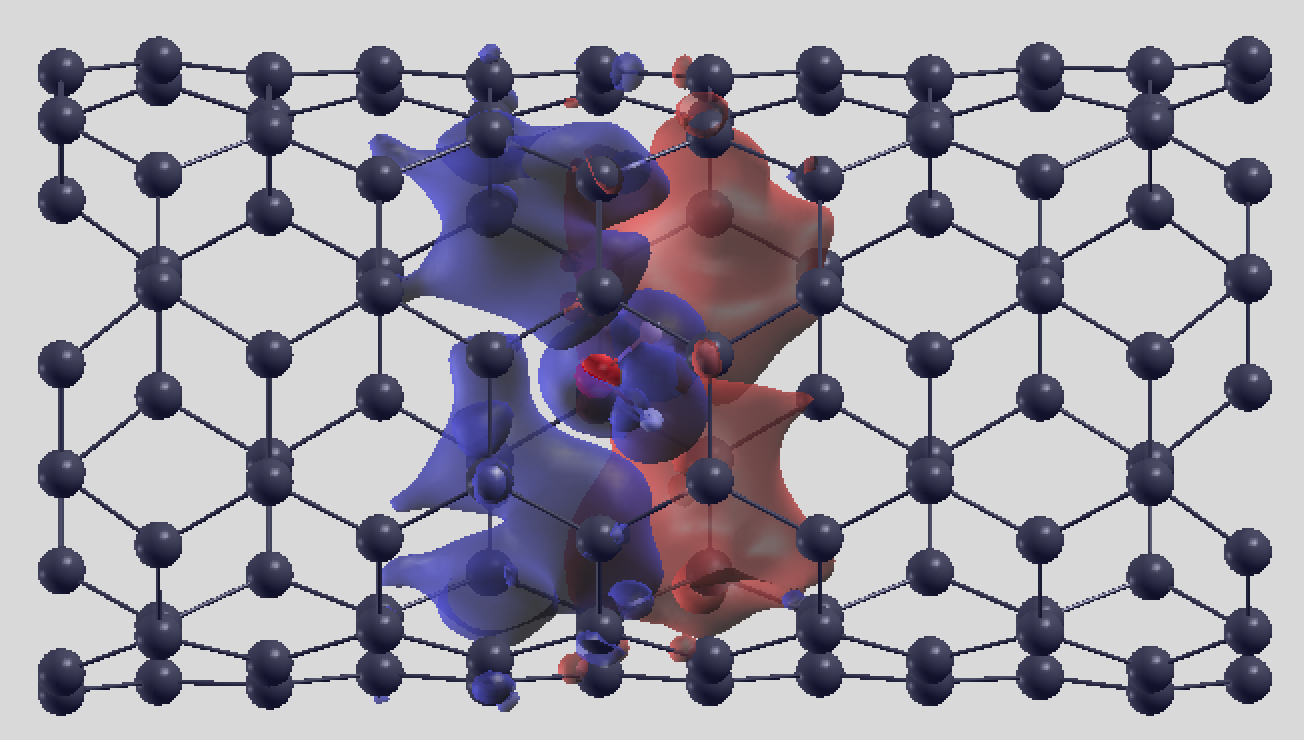}
}
\caption{Differential electron charge density, $\Delta \rho$, 
for H$_2$O@C$_{120}$ (5,5), with isosurfaces shown at 
$\pm 2\times 10^{-3}\, e/{\rm \AA}^3$.
Red areas indicate electron density gain, while blue areas indicate loss
of electron density relative to the empty C$_{120}$ (5,5) nanotube and the 
isolated H$_2$O molecule.} 
\label{fig3}
\end{figure}
                      
\begin{figure}
\centerline{
\includegraphics[width=13cm]{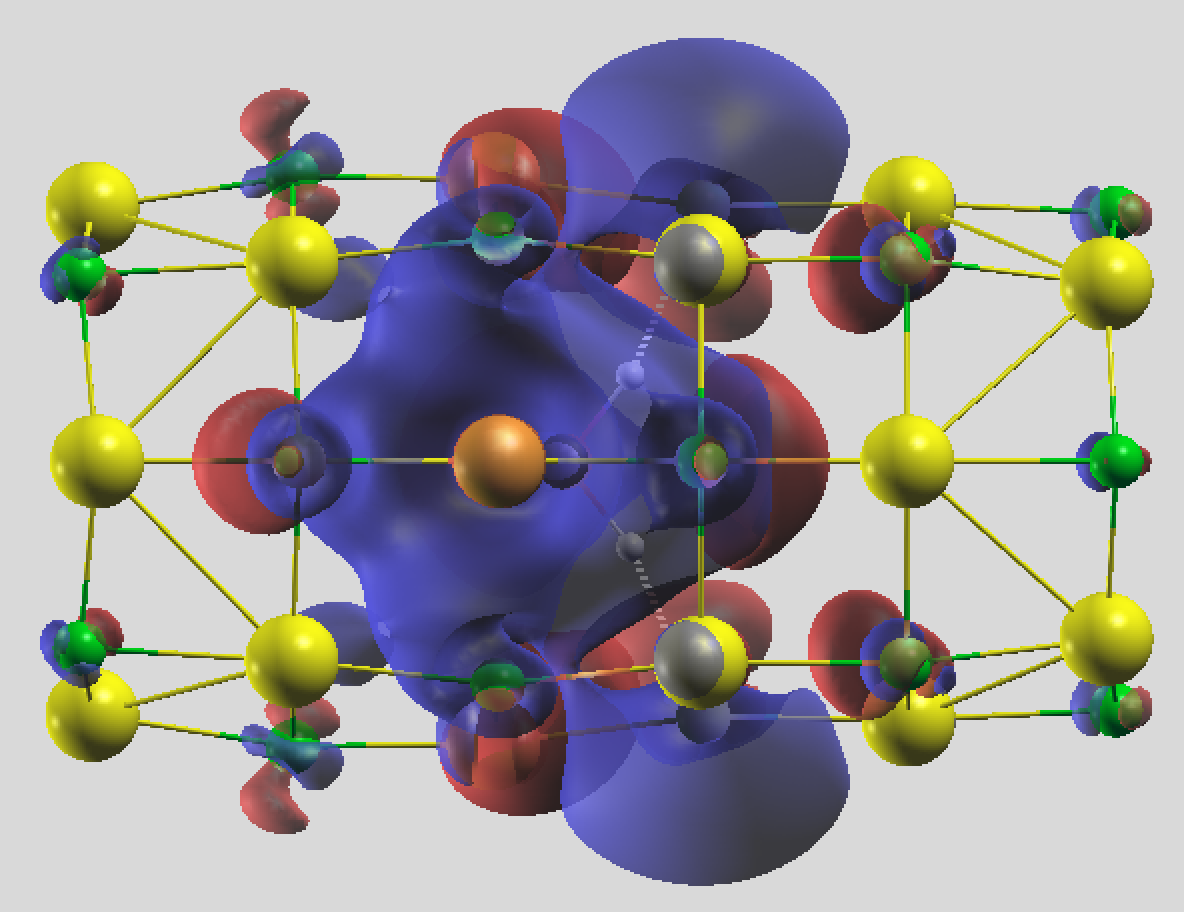}
}
\caption{Differential electron charge density, $\Delta \rho$, 
for H$_2$O@Li$_{24}$F$_{24}${\it oct-6}, with isosurfaces shown at 
$\pm 2\times 10^{-3}\, e/{\rm \AA}^3$.
Red areas indicate electron density gain, while blue areas indicate loss
of electron density relative to the empty Li$_{24}$F$_{24}${\it oct-6} 
nanotube and the isolated H$_2$O molecule.} 
\label{fig4}
\end{figure}
                      
\begin{figure}
\centerline{
\includegraphics[width=13cm]{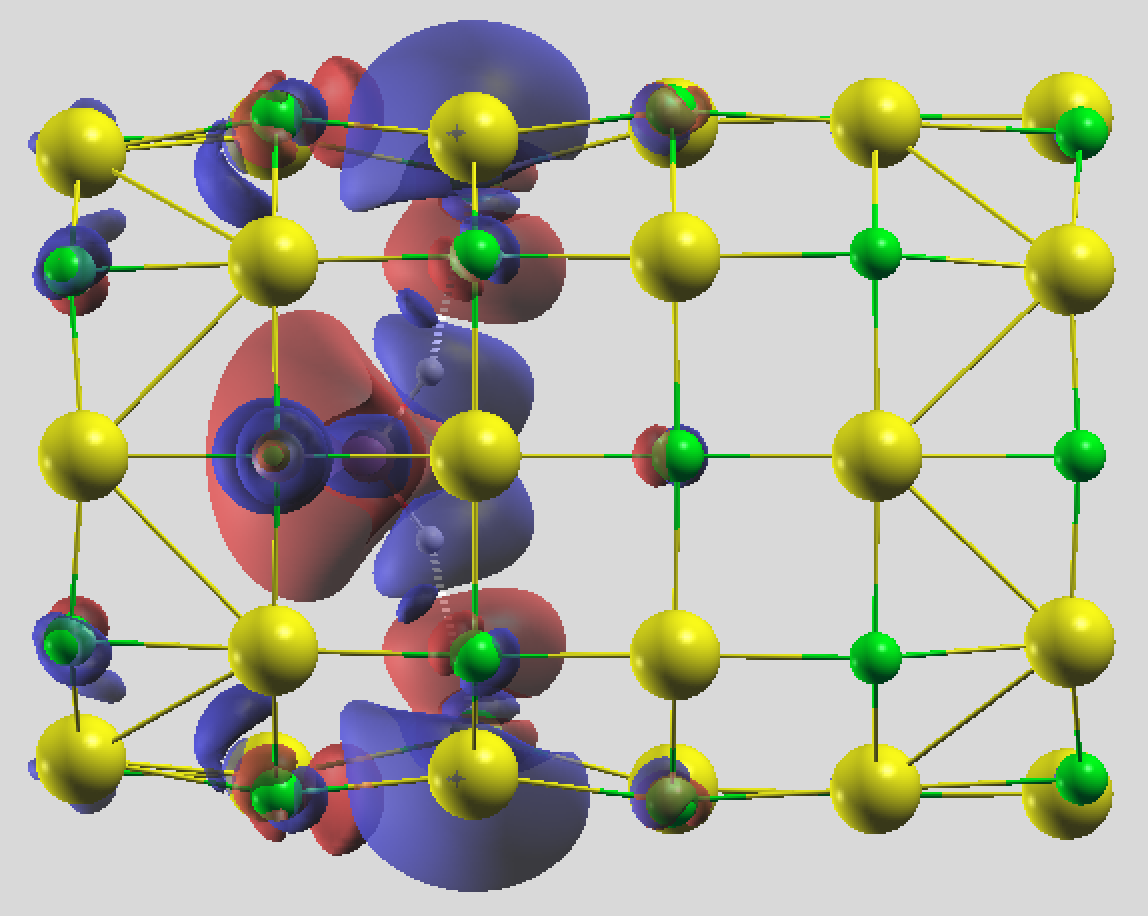}
}
\caption{Differential electron charge density, $\Delta \rho$, 
for H$_2$O@Li$_{36}$F$_{36}${\it dod-6}, with isosurfaces shown at 
$\pm 2\times 10^{-3}\, e/{\rm \AA}^3$.
Red areas indicate electron density gain, while blue areas indicate loss
of electron density relative to the empty Li$_{36}$F$_{36}${\it dod-6} 
nanotube and the isolated H$_2$O molecule.} 
\label{fig5}
\end{figure}
                      
In Figures 6-8 we also plot the one-dimensional profiles $\delta\rho(z)$, 
computed along the nanotube $z$ axis, as a function of $z$-coordinate values, 
by integrating $\delta\rho$ over the corresponding, orthogonal $x$,$y$ planes. 

\begin{figure}
\centerline{
\includegraphics[width=16cm]{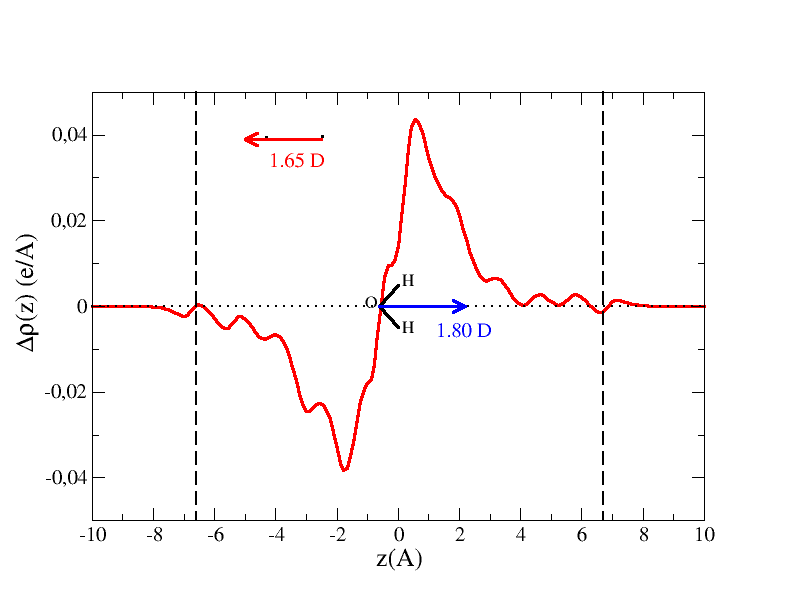}
}
\caption{Differential electron charge density, $\Delta \rho (z)$, along the
nanotube $z$ axis, in H$_2$O@C$_{120}$-(5,5).
The vertical, black, dashed lines indicate the boundaries of the nanotube, 
while the red arrow represents the induced dipole moment with a numerical
value obtained by integration on the $z$ axis (see text). The dipole moment
of the water molecule is also reported in blue.}
\label{fig6}
\end{figure}

\begin{figure}
\centerline{
\includegraphics[width=16cm]{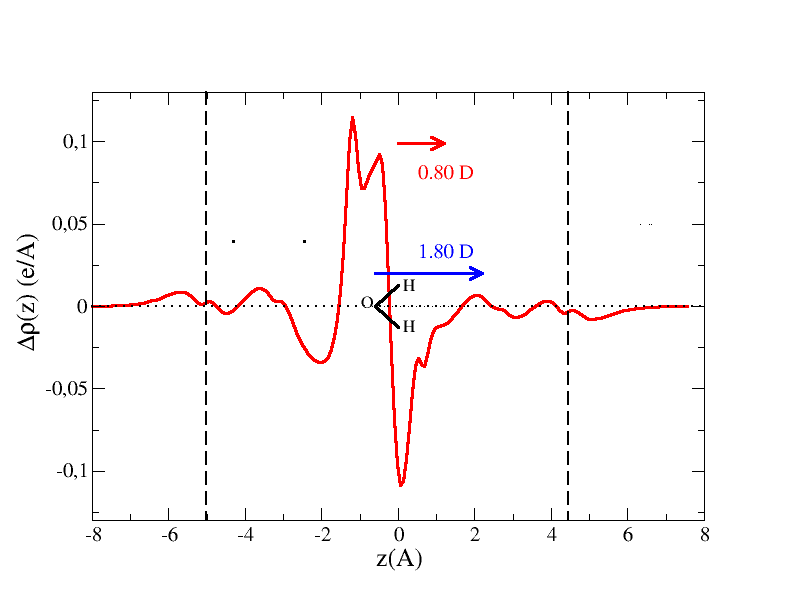}
}
\caption{Differential electron charge density, $\Delta \rho (z)$, along the
nanotube $z$ axis, in H$_2$O@Li$_{24}$F$_{24}${\it oct-6}.
The vertical, black, dashed lines indicate the boundaries of the nanotube, 
while the red arrow represents the induced dipole moment with a numerical
value obtained by integration on the $z$ axis (see text). The dipole moment
of the water molecule is also reported in blue.}
\label{fig7}
\end{figure}

\begin{figure}
\centerline{
\includegraphics[width=16cm]{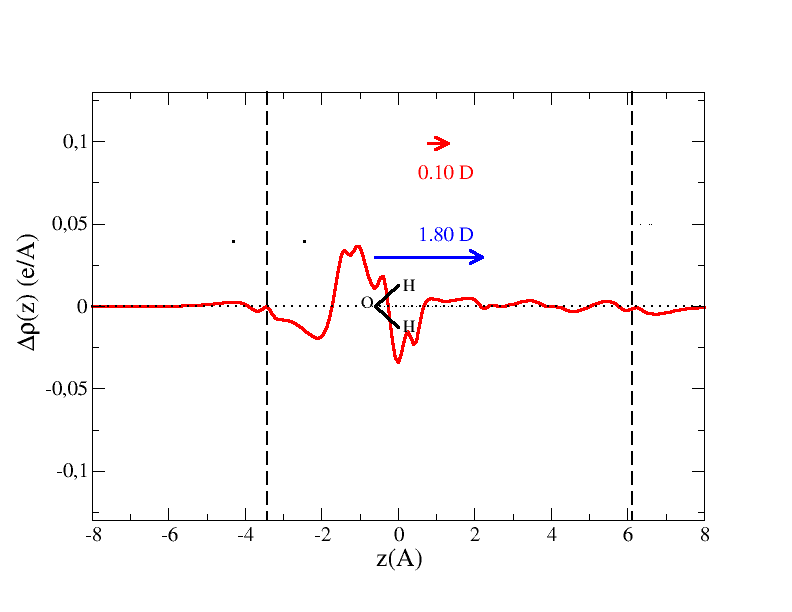}
}
\caption{Differential electron charge density, $\Delta \rho (z)$, along the
nanotube $z$ axis, in H$_2$O@Li$_{36}$F$_{36}${\it dod-6}.
The vertical, black, dashed lines indicate the boundaries of the nanotube, 
while the red arrow represents the induced dipole moment with a numerical
value obtained by integration on the $z$ axis (see text). The dipole moment
of the water molecule is also reported in blue.}
\label{fig8}
\end{figure}

As can be seen in Figure 3, in H$_2$O@C$_{120}$ (5,5) there is a 
pronounced electron charge accumulation in the region 
between the H atoms and the nanotube and a corresponding charge depletion 
around the O atom leading to the formation of the
counteracting dipole moment which considerably reduces the effective dipole 
moment of the endohedral complex; clearly the overall response of the 
C$_{120}$ (5,5)
nanotube to the H$_2$O molecule dipole moment is a significant charge 
density shift. 
One can make more quantitative the information contained in Figure 6 
by evaluating the 
induced dipole moment as: $\mu_{\rm ind} = -\int dz\,z \delta\rho(z)$, where 
$\delta\rho(z)$ has been defined above.
The numerical value of $\mu_{\rm ind}=$1.65 D is found to 
coincide (see Figure 6) with that of $\delta\mu$ reported in Table II.

In ionic nanotubes the situation is more complex.
Considering H$_2$O@Li$_{24}$F$_{24}${\it oct-6}, one can see (Figure 4)
that the regions where
the differential charge density is mostly localized are in the center of the
nanotube, around the O atom of the water molecule, and also in close vicinity of
the 2 H atoms facing 2 F atoms of the nanotube. 
In Figure 7 we plot $\delta\rho(z)$. The corresponding induced dipole moment,
$\mu_{\rm ind}=0.17$ D, is due to the rearranged electron density. 
One can therefore conclude that, for this system, the dipole-amplification 
(antiscreening) effect ($\delta\mu=+0.63$ D, see Table II) 
is due, for about 30\% to the deformation of the electronic cloud, 
while the remaining portion is
due to the relative displacements of positive and negative ions.
In fact the water molecule encapsulated inside the nanotube interacts with 
the positive and negative ions, slightly distorting the nanostructure, with 
the formation of 2 Hydrogen bonds too, as discussed above. 
This effect contributes significantly to the amplified electric dipole moment. 
In order to better characterize these distortions one can consider $z_+$,
that is the z component of the charge center of the positive ions Li$^+$, 
and $z_-$, the z component of the charge center of the negative ions F$^-$; 
then one can evaluate $d\pm = z_+ - z_-$, that is the
difference between the $z$-coordinates of the charge centers of the 
positive and negative ions ($d\pm$ is conveniently expressed in m\AA\ being
a very small quantity). As expected, $d\pm$ increases significantly upon
water-molecule encapsulation (1.7 m\AA) from the value computed in the
isolated Li$_{24}$F$_{24}${\it oct-6} nanotube (0.7 m\AA).
A more detailed description can be obtained (similarly
to what done in ref.\cite{psil23}) by plotting the $z$-coordinate 
changes of the ions, upon encapsulation of
the water molecule, as a function of the $z$ position of the ions themselves. 
Since each transversal section of the Li$_{24}$F$_{24}${\it oct-6} nanotube has 8 
alternating ions (4 Li$^+$ and 4 F$^-$ ions) we consider the average position
along $z$ of each of the two sets of 4 ions: $z_{+i}$ for the 4 Li$^+$ ions 
and $z_{-i}$ for the 4 F$^-$ ions of the $i$-th section. 
Figure 9 shows the behavior of $z_{+i}$ and $z_{-i}$ in the 
Li$_{24}$F$_{24}${\it oct-6} nanotube (the dotted lines are just a guide
to the eyes).

\begin{figure}
\centerline{
\includegraphics[width=16cm]{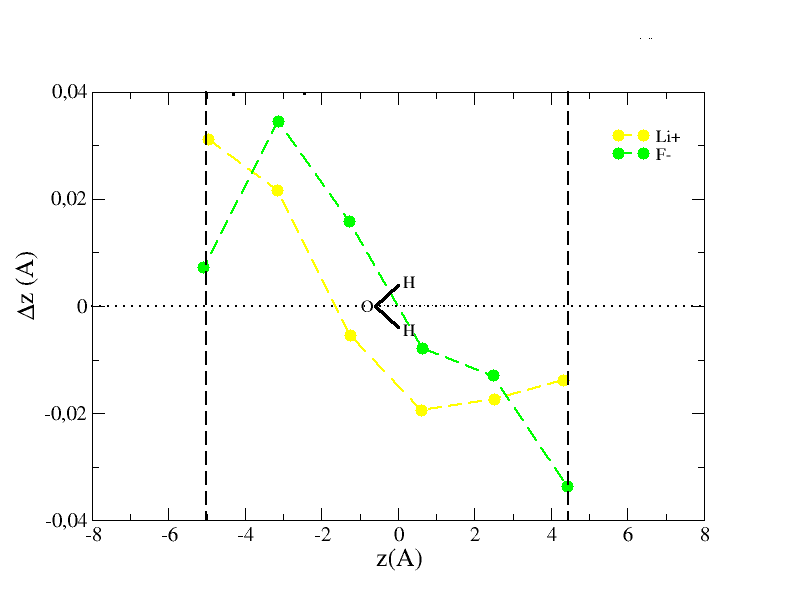}
}
\caption{H$_2$O@Li$_{24}$F$_{24}${\it oct-6} endohedral complex: displacements,
along the nanotube $z$ axis, of Li$^+$ and F$^-$ ions of the
Li$_{24}$F$_{24}${\it oct-6} ionic nanotube upon encapsulation of 
H$_2$O: small circles
represent the actual displacements of the ions, while the dashed lines
are just a guide for the eye.
The vertical, black, dashed lines indicate the positions of the nanotube 
boundaries.} 
\label{fig9}
\end{figure}

Interestingly, lithium-fluoride endohedral structures with the 
dodecagonal section, differently from those with the octagonal one, exhibit
a clear screening effect since the total-dipole moment is reduced 
with respect to that of the isolated molecules (see Table II). 
As can be seen in Figure 2, the endohedral H$_2$O@Li$_{36}$F$_{36}${\it dod-6} 
structure is characterized by a significant deformation in the transversal 
direction (see also the above discussion) with the water molecule that forms
2 Hydrogen bonds with 2 F atoms of the nanotube in this case too. 
The differential charge density $\delta\rho$ of 
H$_2$O@Li$_{36}$F$_{36}${\it dod-6} is
plotted in Figure 5. A clear difference is evident with respect to the 
$\delta\rho$ distribution in H$_2$O@Li$_{24}$F$_{24}${\it oct-6}: 
the electronic charge variations are much smaller (also compare Figure 7
with Figure 8) than those observed in H$_2$O@Li$_{24}$F$_{24}${\it oct-6}.
As a consequence, by integration one obtains a smaller induced dipole
moment (see Figure 8) of 0.10 D. However, in this case, the slight increase
of the dipole moment is more than compensated by the dipole-moment
reduction due to the relative displacements of postive and negative
ions, leading to an averall decrease of the the total dipole moment
(screening effect). As can be seen in Figure 10, the behavior of these
displacements is both qualitatively and quantitatively
different from that of the H$_2$O@Li$_{24}$F$_{24}${\it oct-6} case.
From one hand in H$_2$O@Li$_{24}$F$_{24}${\it oct-6} the displacements 
are almost double than in H$_2$O@Li$_{36}$F$_{36}${\it dod-6}; from the
other in the dodecagonal nanotube case the Li$^+$ ions displacements
are on average smaller than those of the F$^-$ ones, leading to an 
overall distance between positive and negative ion centers along $z$ 
of $d\pm =-4.6$ m\AA. Such a negative value indicates that in 
H$_2$O@Li$_{36}$F$_{36}${\it dod-6} the ion displacements induce a
screening effect, at a difference from H$_2$O@Li$_{24}$F$_{24}${\it oct-6},
where instead the opposite effect is observed.   

\begin{figure}
\centerline{
\includegraphics[width=16cm]{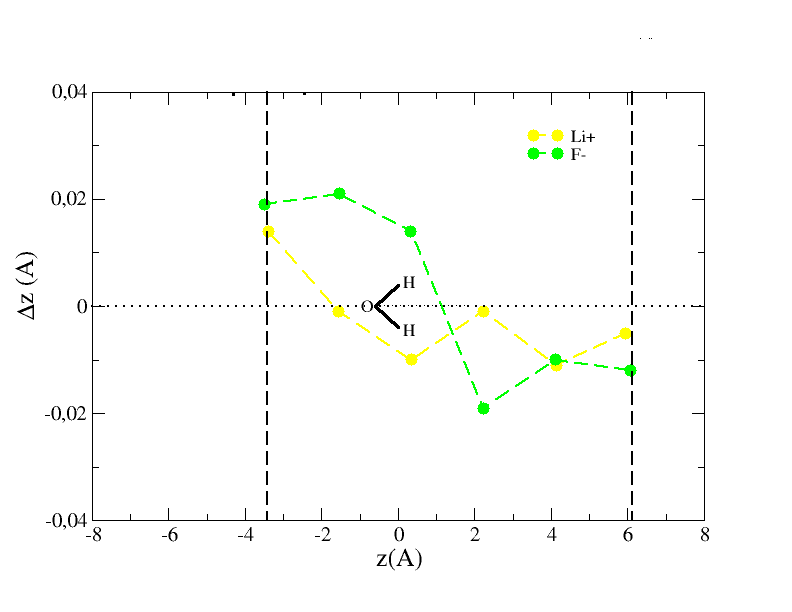}
}
\caption{H$_2$O@Li$_{36}$F$_{36}${\it dod-6} endohedral complex: displacements,
along the nanotube $z$ axis, of Li$^+$ and F$^-$ ions of the
Li$_{36}$F$_{36}${\it oct-6} ionic nanotube upon encapsulation of 
H$_2$O: small circles
represent the actual displacements of the ions, while the dashed lines
are just a guide for the eye.
The vertical, black, dashed lines indicate the positions of the nanotube 
boundaries.} 
\label{fig10}
\end{figure}

The changes in the dipole moment of endohedral LiF nanotubes upon
encapsulation of a molecule can also be semiquantitatively reproduced
by using a simple electrostatic model, where the Li$^+$ and F$^-$ ions of
the nanotubes are replaced by positive and negative charges 
(+0.71 $e$ and -0.71 $e$, respectively, where the effective, partial 
charges were estimated in ref.\cite{psil23}).
The electric field $E$ generated by these charges at the position of the
encapsulated molecule gives rise to an induced dipole moment which
can be easily evaluated as $\Delta\mu = \alpha E$, 
where $\alpha$ is the
polarizability of the encapsulated molecule, which is assumed to be
isotropic (a good approximation for water). Considering the 
induced dipole moment along the nanotube axis one finds that
$\Delta\mu=$ +0.66 and +0.36 D for H$_2$O@Li$_{24}$F$_{24}${\it oct-6}
and HF@Li$_{24}$F$_{24}${\it oct-6}, respectively, in reasonable agreement 
with the corresponding increases of the dipole moments (+0.63 and +0.29 D) 
reported in Table II (antiscreening effect).
Instead, in Li$_{36}$F$_{36}${\it dod-6} the relatively small
induced dipole moments (+0.21 and +0.03 D for encapsulated
H$_2$O and HF molecule) are more than
compensated by the counteracting dipole moment (-0.57 and -0.50 D) 
induced in the nanotube, due to the rearrangement of the Li and F atoms, 
thus leading to an overall reduction of the total dipole moment 
(screening effect). 

We have also verified that, by replacing the rVV10 DFT functional with 
the PBE one (unable to properly describe the vdW effcts) 
the antiscreening and screening behavior 
is qualitatively preserved in Li$_{24}$F$_{24}${\it oct-6} and 
in Li$_{36}$F$_{36}${\it dod-6}, respectively, in spite
of the dramatic reduction of binding of the encapsulated molecules, thus
suggesting that vdW interactions do not play a crucial role in determining
the screening properties, which are mostly due to electrostatic interactions,
as also confirmed by the reasonable description obtained by the 
simple electrostatic model introduced above.

To assess the effect of changing the number of layers in the selected 
ionic nanotubes characterized by octagonal or dodecagonal transversal section, 
we evaluated the dipole moment of the endohedral complexes with 
encapsulated molecules also in nanotubes with one less 
(Li$_{20}$F$_{20}${\it oct-5} and Li$_{30}$F$_{30}${\it dod-5})
and one more layer (Li$_{28}$F$_{28}${\it oct-7} and 
H$_2$O@Li$_{42}$F$_{42}${\it dod-7}) than the 6 layers considered so far.
As can be seen in Table III, the number of layers plays 
a significant role in octagonal LiF nanotubes; in fact the antiscreening 
effect turns out to be maximum for the Li$_{24}$F$_{24}${\it oct-6} 
nanotube with 6 layers.
In the case of the encapsulated water molecule the dipole amplification
is substantially reduced (+15\% vs. +35\%) with 5 layers while the reduction
is less pronounced (+31\% vs. +35\%) with 7 layers. 
For the encapsulated HF molecule the dependence is even stronger: with
5 and 7 layers a screening effect is observed instead of an antiscreening
one, with a decrease of the
dipole moment: from +16\% to -21\% and from +16\% to -6\% with 5 and
7 layers, respectively. 
Instead in dodecagonal LiF nanotubes a milder dependence on the 
number of nanotube layers is found, with the screening behavior 
which is observed in all the considered cases. 

The binding energies of the encapsulated molecules do not vary much with
the number of layers (see Table III), but for the Li$_{42}$F$_{42}${\it dod-7}
nanotube. In fact, in this case $E_{\rm bind}$ is severely reduced (in 
absolute value). We have verified that this is due to a different geometric
structure of the endohedral H$_2$O@Li$_{42}$F$_{42}${\it dod-7} and 
HF@Li$_{42}$F$_{42}${\it dod-7} nanostructures: in fact 
in these cases the dodecagonal nanotube is more rigid and thus much less
distorted upon molecule encapsulation than with 5 or 6 layers, so that the 
interaction with the inside molecule is much reduced due to an average
larger distance from the atoms of the nanotube.

\begin{table}
\caption{Electronic dipole moment (along the nanotube axis) $\mu$ of 
endohedral complexes with H$_2$O or HF molecule inside;
$\delta\mu$ denotes the change of the dipole moment of the endohedral complex 
with respect to those computed for the encapsulated H$_2$O and HF molecules 
when they are isolated (1.80 and 1.79 D, respectively). 
$E_{\rm bind}$ is the binding energy (in square parenthesis using the 
PBE functional
in place of rVV10) of endohedral complexes. {\it oct-6} and {\it dod-6} denote 
octagonal or dodecagonal transversal section with 6 layers.
For comparison we
also report corresponding results obtained considering the H$_2$O and HF molecules
encapsulated in selected nanocages.\cite{psil23}} 
\begin{center}
\begin{tabular}{|c|c|c|c|}
\hline
system & $\mu$ (D) & $\delta\mu$ (D) &  $E_{\rm bind}$ (meV) \\ 
\hline
H$_2$O@C$_{96}$ (4,4)                &  0.07   & -1.74 (-96\%) & +460 \\
HF@C$_{96}$ (4,4)                    &  0.09   & -1.69 (-95\%) & +101 \\
H$_2$O@C$_{120}$ (5,5)               &  0.15   & -1.65 (-92\%) & -442 \\
HF@C$_{120}$ (5,5)                   &  0.14   & -1.64 (-92\%) & -332 \\
\hline
H$_2$O@B$_{48}$N$_{48}$ (4,4)        &  1.20   & -0.60 (-34\%) & -554 \\ 
HF@B$_{48}$N$_{48}$ (4,4)            &  1.06   & -0.72 (-40\%) &  +47 \\
\hline
H$_2$O@Be$_{48}$O$_{48}$ (4,4)       &  1.52   & -0.29 (-16\%) & -195 \\ 
HF@Be$_{48}$O$_{48}$ (4,4)           &  1.12   & -0.66 (-37\%) & -165 \\
\hline
H$_2$O@Ga$_{48}$N$_{48}$ (4,4)       &  0.67   & -1.12 (-63\%) & -290 \\
HF@Ga$_{48}$N$_{48}$ (4,4)           &  0.71   & -1.09 (-60\%) & -403 \\
\hline
H$_2$O@Li$_{24}$F$_{24}${\it oct-6}  &  2.43   & +0.63 (+35\%) & -230[+302] \\
HF@Li$_{24}$F$_{24}${\it oct-6}      &  2.08   & +0.29 (+16\%) & -201[ +67] \\
\hline
H$_2$O@Li$_{36}$F$_{36}${\it dod-6}  &  1.14   & -0.66 (-37\%) & -517[-120] \\
HF@Li$_{36}$F$_{36}${\it dod-6}      &  1.22   & -0.56 (-31\%) & -573[ -59] \\
\hline
\hline
H$_2$O@C$_{60}$  nanocage            &  0.52   & -1.34 (-72\%) & -554[  +9] \\
HF@C$_{60}$  nanocage                &  0.52   & -1.26 (-71\%) & -479[ -64] \\
H$_2$O@B$_{36}$N$_{36}$ nanocage     &  1.17   & -0.69 (-37\%) & -547[ -84] \\
HF@B$_{36}$N$_{36}$ nanocage         &  1.09   & -0.69 (-39\%) & -374[ -63] \\
H$_2$O@Be$_{36}$O$_{36}$ nanocage    &  1.61   & -0.25 (-13\%) & -535[-187] \\
HF@Be$_{36}$O$_{36}$ nanocage        &  1.63   & -0.15 ( -8\%) & -322[ -97] \\
H$_2$O@Li$_{36}$F$_{36}$ nanocage    &  2.20   & +0.34 (+18\%) & -152[ -76] \\
HF@Li$_{36}$F$_{36}$ nanocage        &  2.13   & +0.35 (+20\%) & -100[ -59] \\
\hline
\end{tabular}                                                
\end{center}
\label{table2}                                  
\end{table}

\begin{table}
\caption{Electronic dipole moment (along the nanotube axis) $\mu$ of 
endohedral complexes with H$_2$O or HF molecule inside;
$\delta\mu$ denotes the change of the dipole moment of the endohedral complex 
with respect to those computed for the encapsulated H$_2$O and HF molecules 
when they are isolated (1.80 and 1.79 D, respectively). 
$E_{\rm bind}$ is the binding energy of endohedral complexes. 
{\it oct-x} and {\it dod-x} denote 
octagonal or dodecagonal transversal section with {\it x=5,6,7} layers.}
\begin{center}
\begin{tabular}{|c|c|c|c|}
\hline
system & $\mu$ (D) & $\delta\mu$ (D) &  $E_{\rm bind}$ (meV) \\ 
\hline
H$_2$O@Li$_{20}$F$_{20}${\it oct-5}  &  2.07   & +0.27 (+15\%) & -193 \\
H$_2$O@Li$_{24}$F$_{24}${\it oct-6}  &  2.43   & +0.63 (+35\%) & -230 \\
H$_2$O@Li$_{28}$F$_{28}${\it oct-7}  &  2.36   & +0.56 (+31\%) & -228 \\
HF@Li$_{20}$F$_{20}${\it oct-5}      &  1.40   & -0.38 (-21\%) & -183 \\
HF@Li$_{24}$F$_{24}${\it oct-6}      &  2.08   & +0.29 (+16\%) & -201 \\
HF@Li$_{28}$F$_{28}${\it oct-7}      &  1.68   & -0.11 ( -6\%) & -204 \\
\hline
H$_2$O@Li$_{30}$F$_{30}${\it dod-5}  &  1.37   & -0.43 (-24\%) & -578 \\
H$_2$O@Li$_{36}$F$_{36}${\it dod-6}  &  1.14   & -0.66 (-37\%) & -517 \\
H$_2$O@Li$_{42}$F$_{42}${\it dod-7}  &  1.28   & -0.52 (-29\%) & -265 \\
HF@Li$_{30}$F$_{30}${\it dod-5}      &  1.05   & -0.74 (-41\%) & -592 \\
HF@Li$_{36}$F$_{36}${\it dod-6}      &  1.22   & -0.56 (-31\%) & -573 \\
HF@Li$_{42}$F$_{42}${\it dod-7}      &  1.20   & -0.59 (-33\%) & -168 \\
\hline
\hline
\end{tabular}                                                
\end{center}
\label{table3}                                  
\end{table}

When considering hypothetical endohedral nanostructures it is clearly 
important to have an estimate of the energy cost required (if any) to
encapsulate a given molecule.
To this aim we found the optimal reaction path for encapsulation of a single
H$_2$O or HF molecule, from a position outside to the optimal one inside 
the different considered nanotubes, using the Nudged Elastic Band (NEB) 
method combined with the climbing-image approach and the quasi-Newton
Broyden’s second method.\cite{NEB}

\begin{table}
\caption{Energy barrier (in eV) for the encapsulation of H$_2$O and HF 
molecules in selected nanotubes.}
\begin{center}
\begin{tabular}{|c|c|}
\hline
system & energy barrier (eV) \\ 
\hline
H$_2$O@C$_{96}$ (4,4)                &  0.470 \\
HF@C$_{96}$ (4,4)                    &  0.121 \\
H$_2$O@C$_{120}$ (5,5)               &  0.000 \\
HF@C$_{120}$ (5,5)                   &  0.000 \\
\hline
H$_2$O@B$_{48}$N$_{48}$ (4,4)        &  0.287 \\ 
HF@B$_{48}$N$_{48}$ (4,4)            &  0.062 \\
\hline
H$_2$O@Be$_{48}$O$_{48}$ (4,4)       &  2.573 \\ 
HF@Be$_{48}$O$_{48}$ (4,4)           &  3.733 \\
\hline
H$_2$O@Ga$_{48}$N$_{48}$ (4,4)       &  2.501 \\
HF@Ga$_{48}$N$_{48}$ (4,4)           &  1.703\\
\hline
H$_2$O@Li$_{24}$F$_{24}${\it oct-6}  &  0.499 \\
HF@Li$_{24}$F$_{24}${\it oct-6}      &  0.412 \\
\hline
H$_2$O@Li$_{36}$F$_{36}${\it dod-6}  &  0.000 \\
HF@Li$_{36}$F$_{36}${\it dod-6}      &  0.000 \\
\hline
\end{tabular}                                                
\end{center}
\label{table4}                                  
\end{table}

The energy barriers for the encapsulation of H$_2$O and HF molecules
inside the nanotubes are reported in Table IV.
As can be seen, in case of carbon nanotubes, encapsulation requires 
overcoming an energy barrier of about 0.5 and 0.1 eV in C$_{96}$ (4,4)
for H$_2$O and HF, respectively, while the encapsulation process is
instead barrierless in the larger C$_{120}$ (5,5) nanotube. 
Similarly, in ionic nanotubes the energy barrier is about 0.5 and 0.4 eV in 
Li$_{24}$F$_{24}${\it oct-6} for H$_2$O and HF, respectively, while 
encapsulation is barrierless in Li$_{36}$F$_{36}${\it dod-6}.
In partially ionic nanotubes the energy barriers range from 
about 0.06 and 0.3 eV in B$_{48}$N$_{48}$ (4,4) to much larger
values (2.6 and 3.7 eV) in Be$_{48}$O$_{48}$ (4,4).

Interestingly, it is possible to further amplify the dipole moment
of the endohedral Li$_{24}$F$_{24}${\it oct-6} nanotube by encapsulating 
multiple molecules. For instance we have verified that when a H$_2$O and 
a HF molecule are chained inside the nanotube (Figure 11) the dipole moment
of the endohedral structure is 5.14 D, which is a value considerably
larger than both the sum of the dipole moments of isolated H$_2$O 
and HF molecules (+43\%) and also of the dipole moment of the isolated
H$_2$O-HF complex (+11\%). 

\begin{figure}
\centerline{
\includegraphics[width=12cm]{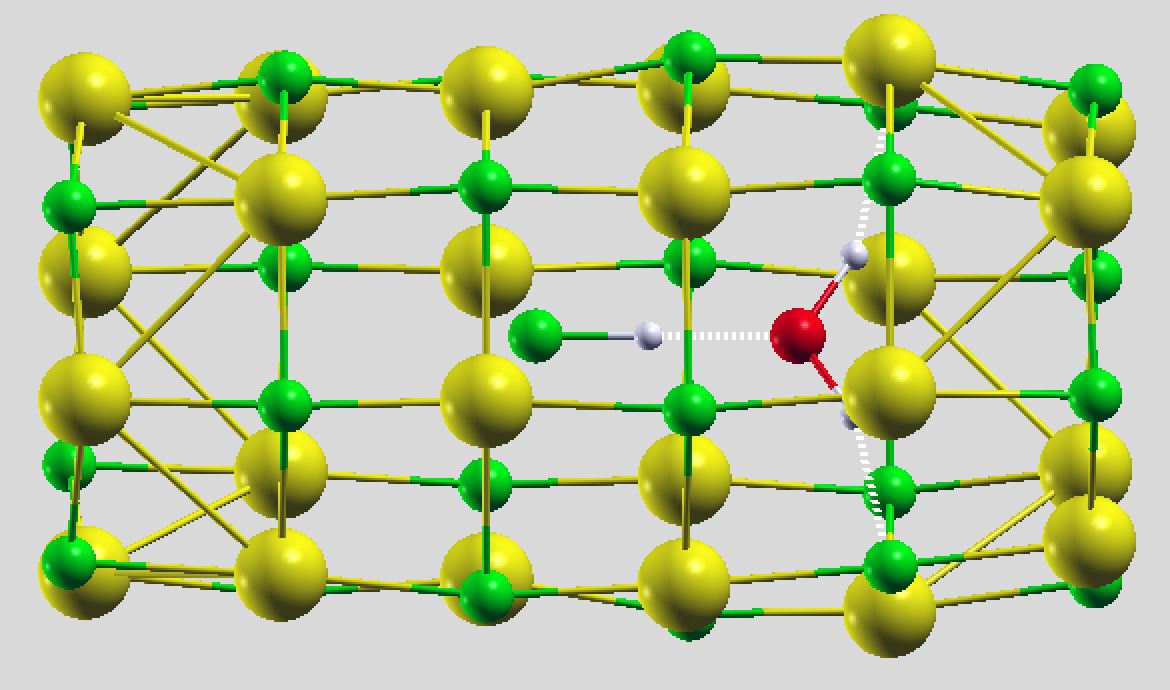}  
}
\caption{HF-H$_2$O@Li$_{24}$F$_{24}${\it oct-6} endohedral nanostructure
(longitudinal view).
Yellow, green, red, and white balls represent Li, F, O ,and H atoms,
respectively. White dashed lines indicate Hydrogen bonds.} 
\label{fig11}
\end{figure}

\section{Conclusions}
We have presented the results of a first-principles study of 
screening effects in endohedral complexes made up of small molecules,
with a finite electronic dipole moment, encapsulated into different 
nanotubes. A detailed analysis of the effective dipole moment of the
complexes and of the electronic charge distribution suggests that
screening effects crucially depend not only on the nature of the
intramolecular bonds but also on the size and the shape 
of the nanotubes and on the specific encapsulated molecule. 
As observed in endohedral nanocages, screening is maximum 
in covalent-bond carbon nanotubes, while it is reduced in partially-ionic 
nanotubes and an antiscreening effect is observed in some ionic nanotubes.
However in this case the scenario is more complex than in corresponding
ionic nanocages.
In fact the specific geometric structure of alkali-halide, ionic 
nanotubes turns out to be crucial for determining the
screening/antiscreening behavior: while nanotubes 
with octagonal transversal section can exhibit an 
antiscreening effect, which quantitatively depends on the number
of layers in the longitudinal direction, instead nanotubes with 
dodecagonal section are always characterized by a reduction of the total 
dipole moment, so that a screening behavior is observed. 
The antiscreening effect in octagonal nanotubes is particularly 
pronounced for the encapsulated water molecule or the HF-H$_2$O chained molecule,
characterized by the formation of Hydrogen bonds.
A detailed structural and electronic analysis has been carried out to
elucidate the occurrence of screening/antiscreening effects in 
different nanotubes.
Our results show that, even in nanotube structures, one can tune the 
dipole moment and generate electrostatic fields at the nanoscale 
without the aid of external potentials. We can also expect some 
transferability of the observed screening/antiscreening effects in 
other nanostructures and 2D materials.
\vfill
\eject

\section{Acknowledgements}
We acknowledge funding from PARD-2024,
relative to the project “Ab initio study of dielectric properties of innovative
nanostructures". Moreover, this work is supported in
part by the MUR Departments of Excellence grant 2023-2027 "Quantum Frontiers"
paradigm for the control of Nanoscale Phenomena”.

\vfill
\eject

\end{document}